\documentclass[10pt]{iopart}
\pdfoutput=1

\expandafter\let\csname equation*\endcsname\relax

\expandafter\let\csname endequation*\endcsname\relax

\usepackage{amsmath}

\usepackage[utf8]{inputenc}
\usepackage{multirow}

\usepackage[dvips]{graphics,graphicx}
\usepackage{url}
\usepackage{subfigure}
\usepackage{etoolbox}
\usepackage{booktabs}
\usepackage[table,xcdraw]{xcolor}
\usepackage{ulem}
\usepackage{overpic}
\usepackage{hyperref}
\usepackage{amssymb}
\usepackage{color}

\newtoggle{draft}

\toggletrue{draft}
\begin{document}

\title[Mechanics of ECC induced alpha particle transport]{Mechanics of ELM control coil induced alpha particle transport}

\author{Konsta S\"arkim\"aki\textsuperscript{1},
  Jari Varje\textsuperscript{1},
  Marina B\'ecoulet\textsuperscript{2},
  Yueqiang Liu\textsuperscript{3},
  Taina Kurki-Suonio\textsuperscript{1}}

\address{\textsuperscript{1}Aalto University, Espoo, Finland}
\address{\textsuperscript{2}CEA, St-Paul-lez-Durance, France}
\address{\textsuperscript{3}General Atomics, San Diego, CA, USA}

\ead{konsta.sarkimaki@aalto.fi}
\vspace{10pt}
\begin{indented}
\item[]\today
\end{indented}

\begin{abstract}\\
Using the orbit-following code ASCOT, we model alpha particle transport in ITER under the influence of ELM control coils (ECCs), toroidal field ripple, and test blanket modules, with emphasis on how the plasma response (PR) modifies the transport mechanisms and fast ion loads on the divertor.
We found that while PR shields the plasma by healing broken flux surfaces, it also opens a new loss channel for marginally trapped particles: PR causes strong toroidal variation of the poloidal field near the X-point which leads to de-localisation of banana tips and collisionless transport.
The reduction in passing particle losses and the increase in marginally trapped particle losses shift divertor loads from targets to the dome and under-the-dome structures.
The plasma response was calculated by both MARS-F and JOREK codes. 
The new transport mechanism was stronger for PR calculated by JOREK which, unlike MARS-F, explicitly includes the X-point.
\end{abstract}

\vspace{2pc}
\noindent{\it Keywords}: ELM control coils, ITER, fast ions, plasma response


%\maketitle
\ioptwocol

\section{Introduction}
\label{sec:Introduction}

ITER will be equipped with ELM control coils (ECCs) to mitigate the detrimental effects of edge localized modes (ELMs).
Their application compromises the axisymmetry of ITER and creates a stochastic field line region at the edge, thus possibly deteriorating the confinement of fast ions such as fusion alphas.
%Their application compromises the axisymmetry of ITER at the edge, thus possibly deteriorating the confinement of fast particles, i.e., fusion alphas as well as NBI and ICRH generated ions in MeV range. 
%Indeed, the first studies of fast ion confinement in the presence of ECCs indicated up to 25\% losses for beam ions, mainly along the edge stochastic field lines. [tanne Shinohara-2012 ja ehkä Tani. Ja Tuomas]. 
%However, in these studies the response of the plasma to the perturbation was ignored, corresponding to vacuum approximation.
Understanding the ECC-induced fast-ion transport has mostly relied on models using the so-called vacuum approximation where the plasma response (PR) has been ignored~\cite{koskela2012iter,tani2011effects,shinohara2012effects}.
Lately, modelling of PR with MHD codes has progressed rapidly~\cite{orain2013non,liu2017comparative}, and ECCs have gained renewed interest in the fast ion community~\cite{shinohara2016investigation,sanchis20163d,akers20163d,garcia2016role,pfefferle2014impact}.
%In the vacuum approximation fast ion losses arise mainly from particles following stochastic field lines~\cite{shinohara2011effects}, PR has been found to reduce the stochasticity by healing some of the broken flux surfaces.
Plasma response 'heals' part of the stochastic region and, thus, the vacuum approximation is expected to give a conservative estimate on total losses. 
This was indeed the case when losses in vacuum approximation were compared to losses including PR for ITER~\cite{varje2016effect}.
However, PR was found to redistribute the loads on the divertor: dome and under-the-dome structure received higher loads than with the vacuum approximation, while targets received lower loads.
Furthermore, the ions were lost at higher energy.
The redistribution is unfortunate since the under-the-dome structure holds cooling pipes whose maximum allowed heat load is much smaller than that of other divertor components~\cite{oikawa2012effects}.
%Therefore, vacuum approximation does not give the highest heat loads on the most sensitive components.

To understand why the redistribution of losses occur, and whether this is true in general, we continue the work reported in Ref.~\cite{varje2016effect} by identifying different loss mechanisms.
We model alpha particle transport in the presence of ECCs and PR, with an orbit-following code.
Because ITER-relevant ECC-induced fast-ion simulations involving PR are a relatively new topic, we seek to verify the results by comparing the fast ion transport when PR is provided by two different MHD codes.

\section{Overview of the models used to evaluate the plasma response and fast ion losses}
\label{sec:Tools}

The slowing-down simulations were carried using same methods as in Ref.~\cite{akaslompolo2015iter}.
Here we simulated markers representing alpha particle population until they either thermalized or came in contact with a material surface.
The simulations were carried with the orbit-following code ASCOT~\cite{hirvijoki2014ascot}. 
ASCOT supports an arbitrary 3D magnetic field, given on a cylindrical $Rz\phi$-grid, and as such it can model transport processes arising from a stationary non-axisymmetric magnetic field in addition to neoclassical transport.

The simulations correspond to the ITER baseline $I_p = 15$ MA H-mode scenario where ECCs are operated with $n=3$, $I=45$ kAt.
To obtain PR, the \emph{vacuum field} is decomposed into toroidal harmonics, and the $n=1-6$ harmonics are given to a MHD code.
The results of the MHD code replace the original $n=1-6$ harmonics in the vacuum field and give the full \emph{plasma response field}.

Two different MHD codes were used to calculate the plasma response: MARS-F~\cite{liu2000feedback,liu2016modelling} is a one-fluid, linear (each $n$ solved independently) code solving the full resistive MHD equations in toroidal geometry.
JOREK~\cite{orain2013non} is a two-fluid, non-linear code solving the reduced MHD equations in toroidal geometry which (unlike for MARS-F) includes the divertor and X-point.
JOREK recalculates equilibrium, which is used in corresponding ASCOT simulations.
For ECCs, the poloidal phases, $[\Phi_U, \Phi_M, \Phi_L]$, for upper, middle, and lower coils were taken from Ref.~\cite{evans20133d}: $[86^\circ, 0^\circ, 34^\circ]$.
However, different phases were used in JOREK: $[58^\circ, 0^\circ, 6^\circ]$. % and in MARS-F and vacuum approximation: $[86^\circ, 0^\circ, 34^\circ]$.
As we shall report later on, this difference in coil phases does not play a significant role on the conclusions of this work.
Another difference was the resistivity, which is the main factor in screening.
In JOREK, $S = 7\times10^9$ (central Lundquist number) was used, and $S = 1\times10^{11}$ in MARS-F.

\section{Modification of the loss channels due to ECCs}
\label{sec:Loss channels}

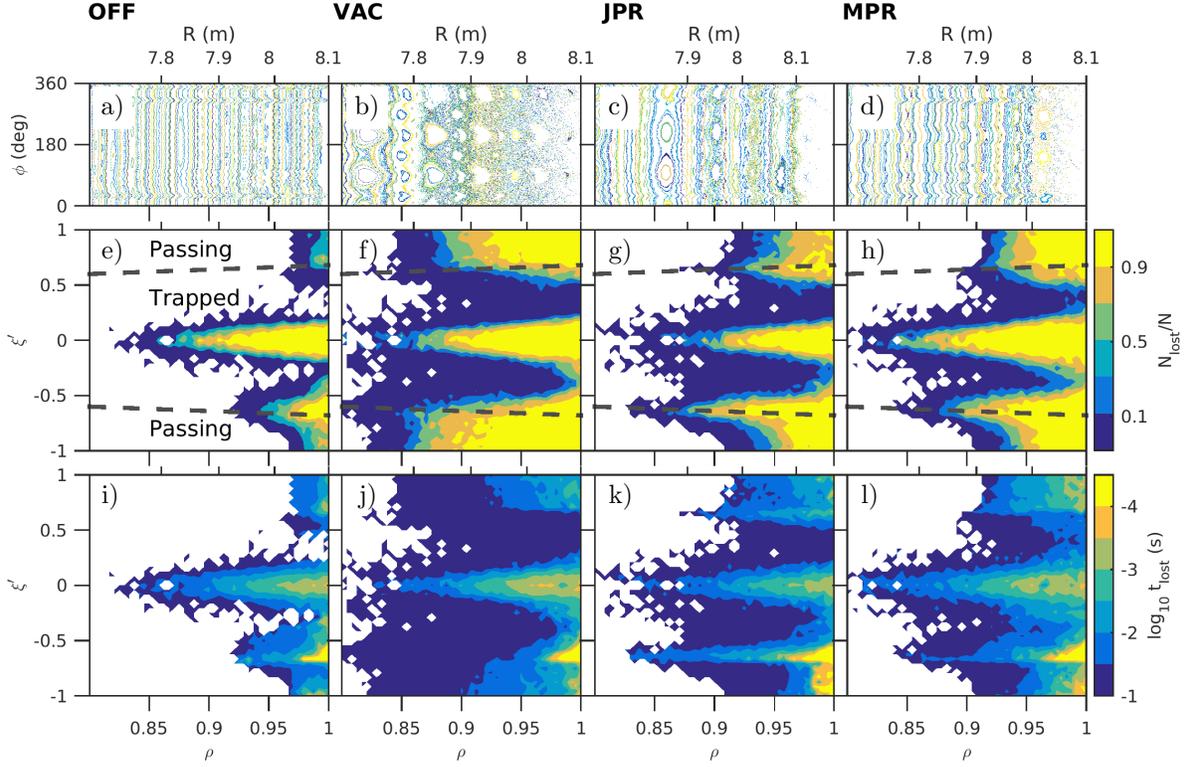
\begin{figure*}[!t]
\centering
\begin{overpic}[width=0.95\textwidth]{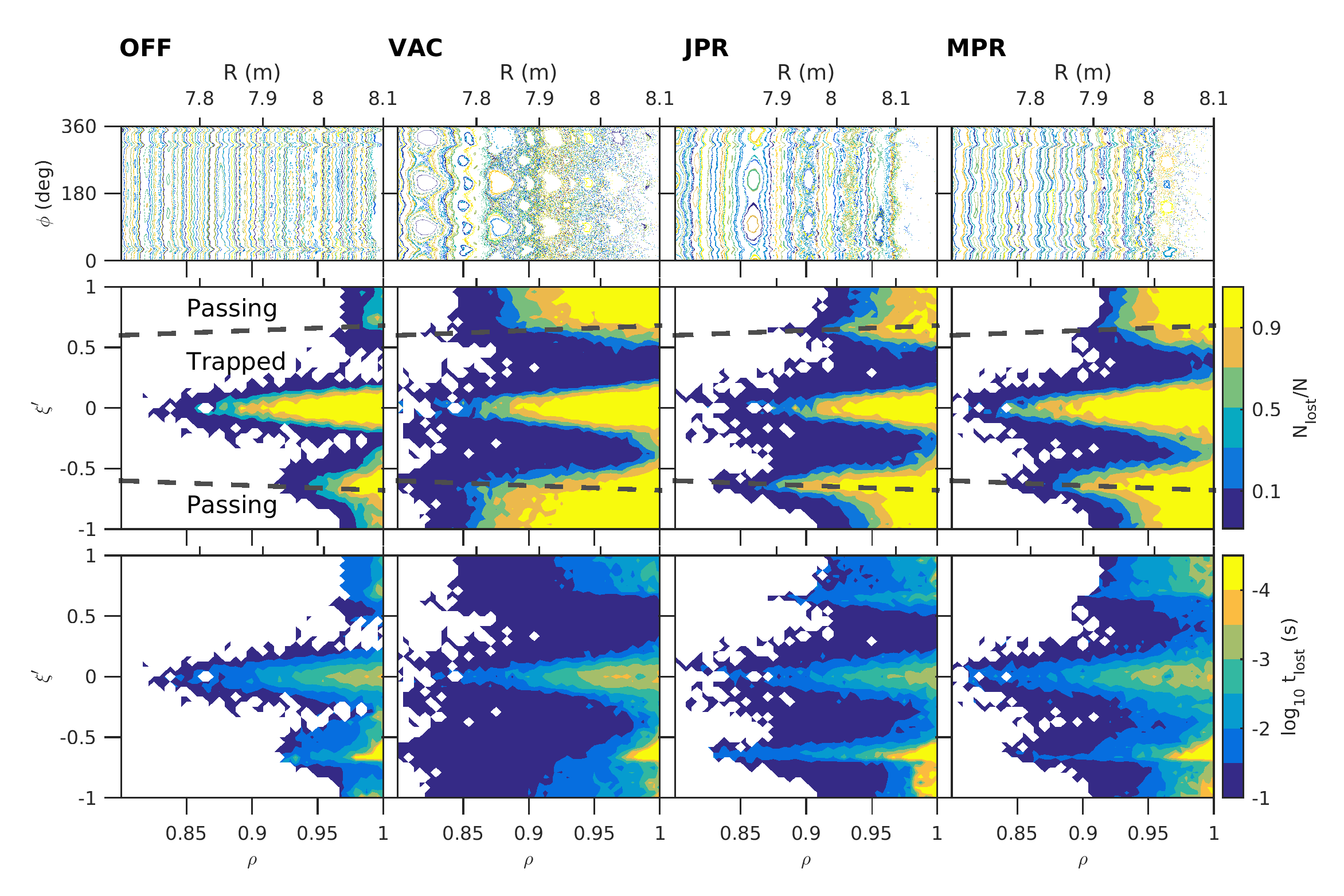}
\put(9.3,54.8){\colorbox{white}{a)}}
\put(30,54.8){\colorbox{white}{b)}}
\put(50.7,54.8){\colorbox{white}{c)}}
\put(71.3,54.8){\colorbox{white}{d)}}
\put(10,43){e)}
\put(31,43){f)}
\put(51.4,43){g)}
\put(72,43){h)}
\put(10,23){i)}
\put(31,23){j)}
\put(51.4,23){k)}
\put(72,23){l)}
\end{overpic}
\caption{Alpha loss dynamics for the different cases. 
(a) - (d) Field-line Poincar\'e-plot at OMP where colors indicate different field lines. Here, $R$ is the major radius corresponding to $\rho$ at OMP.
(e) - (h) Fraction of particles lost, $N_\mathrm{lost}/N$, where $N=N(\rho,\xi')$ is the number of particles whose initial location corresponds to $(\rho,\xi')$.
White regions correspond to no losses.
The dashed lines roughly show the trapped-passing boundary.
(i) - (l) Mean loss time as a function of initial position.}
\label{fig:channels}
\end{figure*}

Slowing-down simulations for alpha particles were done for four cases: no ECCs (\textbf{OFF}), ECCs included in vacuum approximation (\textbf{VAC}), and ECCs included with plasma response solved with JOREK (\textbf{JPR}) or MARS-F (\textbf{MPR}).
All cases include also TF ripple mitigated by ferritic inserts and the test blanket modules (TBMs).
The magnetic field structure in each case is illustrated with a Poincar\'e-plot in the top row in Fig.~\ref{fig:channels}.
Generally, the field structure can be divided into three domains: closed field-line region with healthy flux surfaces, stochastic region where field-lines are chaotic, and laminar region where field-lines intersect wall within a few poloidal orbits and as such appear as sparse or empty areas in Poincar\'e-plots.
In VAC, the coils make the field stochastic down to $\rho \approx 0.85-0.9$ where $\rho$ is the square root of the normalized poloidal flux ($\rho = 1$ at the separatrix).
The plasma response heals some of the flux surfaces but creates a laminar region next to the separatrix.
With JPR, the laminar region extends down to $\rho \approx 0.97$, followed by a narrow stochastic region.
With MPR, the laminar region extends down to $\rho \approx 0.98$ and the stochastic region to $\rho \approx 0.95$.

One would expect that the depth of stochastic/laminar region correlates with passing particle losses.
To see whether this is the case, we map the initial ($R,z,\phi,\xi=v_{||}/v$) position of the simulated markers to a ($\rho, \xi'$)-space where $\rho = \rho(R,z)$ and $\xi'$ is the outer midplane (OMP) pitch value.
The ($\rho$, $\xi'$) coordinates determine not only which particles are trapped but they are also a major factor in determining how a particle responds to the 3D field structure.
We make use of the latter point in the middle row of Fig.~\ref{fig:channels}, where the fraction of lost particles, evaluated as a function of ($\rho, \xi'$), allows us to identify various loss channels (yellow):
\begin{itemize}
\item $\boldsymbol{|}\boldsymbol{\xi'}\boldsymbol{|} \boldsymbol{\lesssim} \mathbf{0.2}.$ 
\emph{Direct-ripple} and \emph{ripple-enhanced} losses. 
Because ECCs only slightly increase magnitude of the TF ripple, this loss channel is similar in each case.
In JPR, the losses are slightly smaller because the outer gap is somewhat larger in JOREK-calculated equilibrium .
\item $\mathbf{0.2} \boldsymbol{\lesssim} \boldsymbol{|}\boldsymbol{\xi'}\boldsymbol{|} \boldsymbol{\lesssim} \mathbf{0.6}.$ 
\emph{Stochastic ripple losses} are also caused by the TF ripple and, hence, are only slightly increased by the ECCs.
No increase was expected between VAC and JPR cases due to JOREK using the reduced MHD model, while there is an increase between VAC and MPR.
\item $\mathbf{0.6} \boldsymbol{\lesssim} \boldsymbol{|}\boldsymbol{\xi'}\boldsymbol{|} \boldsymbol{\lesssim} \mathbf{0.7}.$ 
Marginally trapped particles lost via yet unidentified mechanism(s).
These are present in all cases, enhanced in VAC, and further increased in JPR and MPR.
\item $\mathbf{0.7} \boldsymbol{\lesssim} \boldsymbol{|}\boldsymbol{\xi'}\boldsymbol{|}.$ 
\emph{Stochastic or open field-line losses.}
Nearly all passing particles are lost in regions where the field is stochastic or laminar.
\end{itemize}
The fraction of losses is asymmetric in $\xi'$ because ones with $\xi' < 0$ are born in counter-current direction and, thus, are moved outward by the gradient drift.
%on the inner side of their banana orbit, i.e., on counter-passing trajectories, while converse is true for $\xi' > 0$.

The bottom row of Fig.~\ref{fig:channels} shows the time during which particles were lost.
Bounce time is roughly $10^{-5}$ s, and so \emph{prompt losses} correspond to roughly $t_\mathrm{lost} < 10^{-4}$ s. This is seen to correspond to parameter regime $\xi' \approx 0.6-0.7$, $\rho \gtrsim 0.96$. These are marginally trapped particles whose orbits are wide enough for the particle to exit the plasma and cross material surface.
Dark blue regions in the middle and bottom rows in Fig.~\ref{fig:channels} correspond to slowed-down particles that were not born inside a loss channel but have entered it collisionally.

The results show that VAC indeed gives a conservative estimate on losses but only with respect to passing particles. 
However, the ``spikes" around $\xi'\approx\pm 0.6$ in Figs.~\ref{fig:channels} (g), (h), (k) and (l) shows, once we neglect the prompt losses, that PR gives birth to a new loss channel affecting marginally passing particles.
For convenience, we refer this new mechanism as \emph{perturbed banana transport} (PBT).
PBT features characteristics similar to the ripple transport: it penetrates into the plasma and causes rapid losses which indicates that PBT might also be a convective or resonant process.
As for comparing different PRs, the passing particle transport is stronger in MPR, but PBT is more prominent in JPR.
The larger magnetic islands in JPR does not seem to have a significant effect on transport.

\begin{figure*}[!ht]
\centering
\begin{overpic}[width=0.95\textwidth]{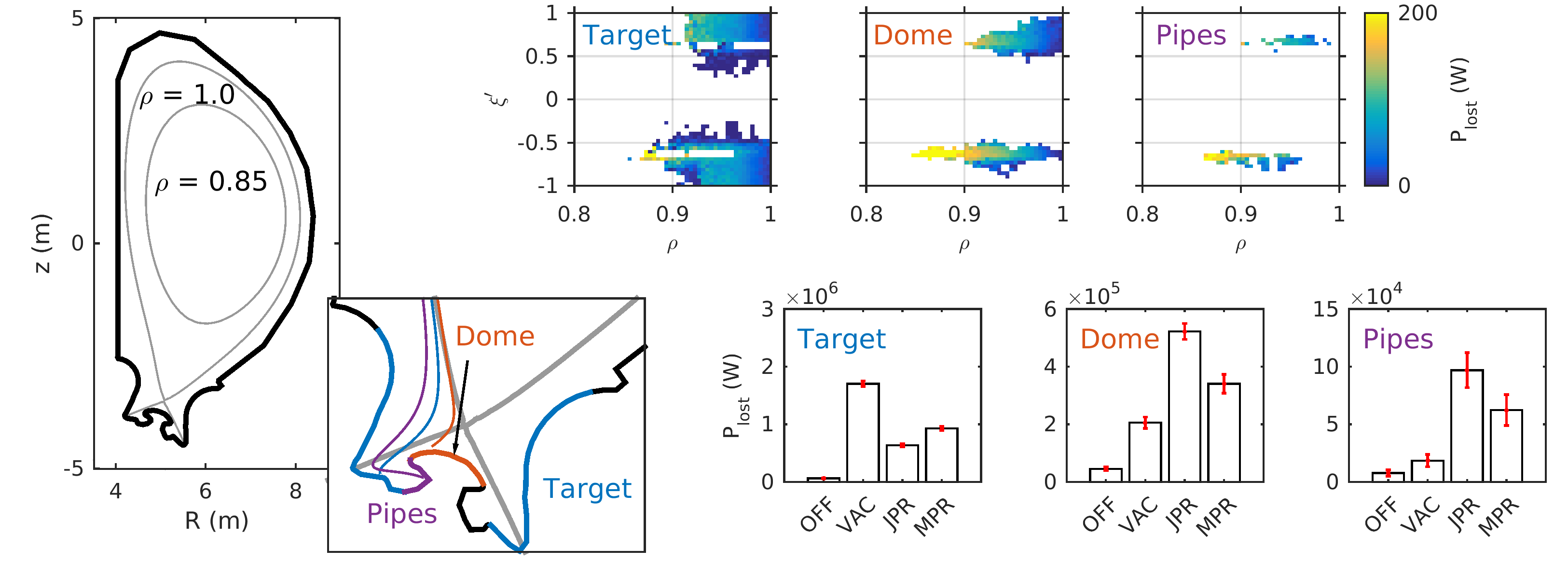}
\put(18,32){a)}
\put(22,12){b)}
\put(31,34){c)}
\put(52,34){d)}
\put(70,34){e)}
\put(46,17){f)}
\put(64,17){g)}
\put(81,17){h)}
\end{overpic}
\caption{Origin of the lost alpha power, $P_\mathrm{lost}$, to the divertor.
Illustration of (a) the first wall and (b) the divertor components: target (blue), dome (orange), and under the dome cooling pipes (purple). 
In (a), the poloidal flux contours shows the region where majority of the losses originate in $Rz$-plane.
The solid trajectories in (b) illustrate how particles end up to the different components.
The origin of the lost power, in $(\rho,\xi')$-space, is shown in (c) - (e).
Note the white horizontal stripes in (c) are due to all particles being lost on or under the dome.
Power load to different divertor components is given in (f) - (h), with red bars indicating the Monte Carlo error.}
\label{fig:divertor}
\end{figure*}
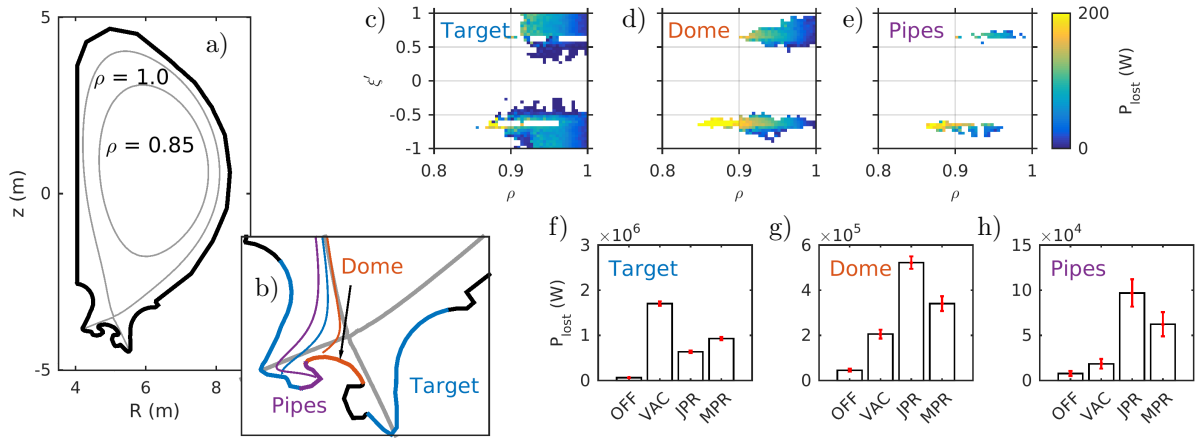

\section{Re-distribution of the divertor loads}
\label{sec:Loss localisations}

Figure~\ref{fig:divertor} shows how the losses are distributed among three different divertor subregions: \emph{target} consisting of inner and outer target plates, \emph{dome umbrella}, and the unprotected \emph{under-the-dome cooling pipes}. %that can withstand much lower heat loads than the other two regions.
In the absence of ECCs, the loads are insignificant. 
Introducing the ECCs in vacuum approximation increases loads on all components.
Including PR decreases the loads on target but increases dome and under-the-dome loads.
The total alpha particle losses were OFF: 0.2 MW, VAC: 2.0 MW, JPR: 1.2 MW, and MPR: 1.5 MW.

To understand the shift in divertor loads, one must first identify the type of particles contributing to it.
The target loads are mainly due to passing particles, but also marginally trapped particles with trajectories falling below the X-point can contribute.
However, losses of marginally trapped particles are predominantly on the dome.
Particles that miss the dome but encounter a second reflection before reaching the target can end up on the pipes.
The divertor loads are shifted because PR decreases passing particle transport while increasing the marginally trapped particle transport.
The dome and under-the-dome loads are higher in JPR compared to MPR because PBT was stronger in the former.

The heat loads to first wall are not shown in Fig.~\ref{fig:divertor} but also they see an increase when PR is included.
%Some of the marginally trapped particles are lost to the wall but PR also enhanced ripple-related losses that contribute to the wall loads.
The heat loads are not worrisome, but this information could be useful when designing fast ion loss detectors.

\section{Mechanism responsible for the marginally trapped particle losses}
\label{sec:Mechanism}

\begin{figure}[h]
\centering
\begin{overpic}[width=0.45\textwidth]{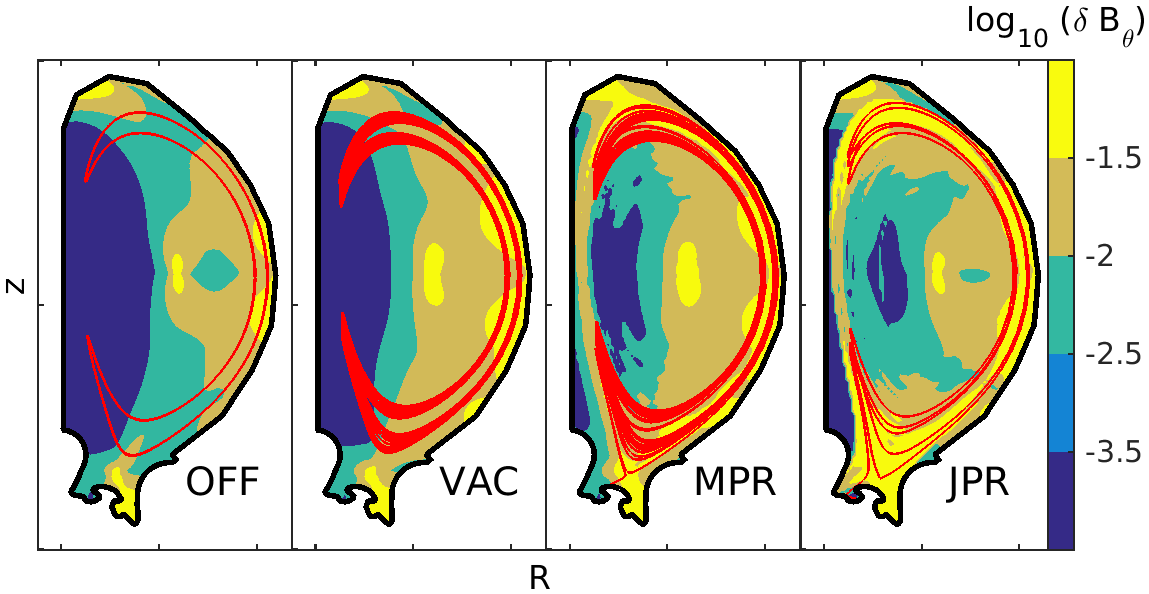}
\put(20,42){a)}
\put(41,42){b)}
\put(64,42){c)}
\put(85,42){d)}
\end{overpic}
\caption{Poloidal field, $B_\theta$, variation along the toroidal direction in different cases.
The contours show the normalized variation, $\delta B_\theta = \left(\left<B_\theta\right>_\mathrm{max}-\left<B_\theta\right>_\mathrm{min}\right) / \left(\left<B_\theta\right>_\mathrm{max}+\left<B_\theta\right>_\mathrm{min}\right)$ where the brackets indicate toroidal maximum or minimum, on an $Rz$-plane.
The red curves shows the trajectory of a sample 3.5 MeV alpha marker with initial location $(\rho = 0.9$, $\xi' = -0.65)$.
The marker was simulated for $1\times10^{-3}$ s with collisions disabled.}
\label{fig:mechanism}
\end{figure}

The mechanisms behind PBT transport appears to be due to the toroidal variation of the poloidal magnetic field near the X-point.
Figure~\ref{fig:mechanism} shows how PR both enlarges the area of significant poloidal field variation and strengthens it locally, in particular near the X-point. 
The same marker alpha, chosen from the $(\rho,\xi')$ region affected by PBT, is simulated for each case and the resulting trajectories are also shown in Fig.~\ref{fig:mechanism}.
The marker shows no radial transport in the OFF case, eliminating the TF ripple as a possible mechanism.
In VAC, the banana tips wander, but only in MPR and JPR these changes are strong enough for the marker to exit the plasma without the aid of collisions.
The changes in the banana tip location is strongest in JPR which also has the strongest poloidal field variation.

The observed behaviour can be explained by noting that the poloidal field strength near the X-point has a large effect on how much $\nabla B$-drift contributes to the orbit width.
If the poloidal field is weak, the particle travels longer almost toroidally. % while converse is true for a strong poloidal field.
The poloidal field around the X-point is small and so even a weak perturbation there has a noticeable effect on the particle trajectory.
When the poloidal field varies toroidally, banana particle precession along the torus means the banana width is not same along subsequent orbits which leads to transport.

We note that this mechanism can be convective, and there might be instances where particles resonate with the perturbation.
However, the reason why PBT is prevalent for alphas with $|\xi'|\approx 0.7$ is due to orbit topology (and not due to a resonance): with smaller $|\xi'|$ the trajectories does not extend beyond the X-point, while for higher $|\xi'|$ values the alphas are already on passing orbits.

%However, the observed transport is not due to particles fulfilling a resonance condition $n\omega_p=k\omega_b$, where $\omega_p$ is the precession frequency, $\omega_b$ is the bounce frequency and $k$ is integer~\cite{poli2008observation}, because such resonant transport is caused by magnetic islands.
%Here, magnetic islands are present only in VAC, while PBT is observed only with plasma response.

%the banana tips points before the X-point while those with larger values are no longer trapped.

\section{Conclusions}
\label{sec:Conclusions}

The inclusion of plasma response to ECCs is essential for an accurate modelling of fast ion transport mechanisms and power loads. 
Assuming that the vacuum approximation would give a conservative estimate on power loads can be misleading. 
Even though the plasma response reduces the \emph{total} power load, it shifts the distribution of the loads, in this case to unprotected components.
We have identified that the shift is due to a new transport mechanism.
Whether resulting heat loads are detrimental to the unprotected components, is a question best answered by codes including accurate divertor geometry~\cite{akers20163d}.
%Vacuum approximation was found not to give conservative estimates on the heat loads on vulnerable components.
%Although the vacuum approximation here resulted in the highest total losses, this might not always be the case because plasma response introduces a completely new transport mechanism.

The reduction in the total losses is mainly due to plasma response healing the flux surfaces and, thus, preventing the passing particle losses. 
On the contrary, plasma response increases the losses of marginally trapped particles, which results in increased power to the cooling pipes under the dome. 
The physical mechanism was identified to result from the toroidal variation of the poloidal field strength, in particular near the X-point.
These main features were present irrespective of which MHD code was used to evaluate the plasma response, but they were more prominent when JOREK was used.
The higher transport of marginally trapped particles seen with JOREK-based plasma response can be attributed to the fact that JOREK model contains X-point geometry.
Therefore, although different coil phases were used here in MARS-F and JOREK, we expect that this finding holds in general.

%However, ECCs in these simulations had different poloidal phases which means the comparison might not have been completely valid.
%Furthermore, the simulations done for this work dismissed the pump-out effect, that is, we did not take into account that the laminar region at the edge also affects the bulk plasma.
%Future work is to address these caveats as well as study beam confinement.

\ack
The work was funded by the Academy of Finland project No. 298126.
Some of the simulations performed for this work were carried out using the computer resources within the Aalto University School of Science `Science-IT' project. 

%\appendix

%\section{Appendix A}

\section*{References}
\bibliographystyle{iopart-num}
\bibliography{rmp_mechanism}

\end{document}